\documentclass[final,3p,times,twocolumn]{elsarticle}
\usepackage{amsmath}
\usepackage[separate-uncertainty]{siunitx} 
\usepackage{cleveref}
\usepackage[acronym,toc,nosuper,nohypertypes={acronym,notation}]{glossaries}
\usepackage[hyphens]{url}
\newacronym{SiLab}{SiLab}{Silizium Labor Bonn}
\newacronym{HISKP}{HISKP}{Helmholtz Institut f\"ur Strahlen- und Kernphysik}
\newacronym{PI}{PI}{Physikalisches Institut}
\newacronym{KIT}{KIT}{Karlsruher Institut f\"ur Technologie}
\newacronym{NIEL}{NIEL}{Non-Ionizing Energy Loss}
\newacronym{IEL}{IEL}{Ionizing Energy Loss}
\newacronym{LHC}{LHC}{Large Hadron Collider}
\newacronym{HL-LHC}{HL-LHC}{High-Luminosity Large Hadron Collider}

\newacronym{CERN}{CERN}{Conseil Européen pour la Recherche Nucléaire}
\newacronym{SEM}{SEM}{Secondary Electron Monitor}
\newacronym{ROE}{R/O Electronics}{Readout Electronics}
\newacronym{ECR}{ECR}{Electron Cyclotron Resonance}
\newacronym{NIST}{NIST}{National Institute of Standards and Technology}
\newacronym{DAQ}{DAQ}{Data Acquisition}
\newacronym{IC}{IC}{Integrated Circuit}
\newacronym{HV}{HV}{High-Voltage}
\newacronym{SMU}{SMU}{Source Measure Unit}
\newacronym{ADC}{ADC}{Analog-to-Digital Converter}
\newacronym{TID}{TID}{Total Ionizing Dose}
\newacronym{FWHM}{FWHM}{Full Width Half Maximum}
\newacronym{PCB}{PCB}{Printed Circuit Board}
\newacronym{SCC}{SCC}{Single Chip Card}
\newacronym{StrlSchG}{StrlSchG}{Strahlenschutzgesetz}
\newacronym{StrlSchV}{StrlSchV}{Strahlenschutzverordnung}
\newacronym{RF}{RF}{Radio Frequency}
\newacronym{RPi}{RPi}{Raspberry Pi}
\newacronym{GUI}{GUI}{Graphical User Interface}
\newacronym{HEP}{HEP}{High Energy Physics}
\newacronym{SEY}{SEY}{Secondary Electron Yield}
\newacronym{LF}{LF}{LFoundry}
\newacronym{BIC}{BIC}{Bonn Isochronous Cyclotron}
\newacronym{SEE}{SEE}{Secondary Electron Emission}
\newacronym{CAD}{CAD}{Computer-Aided Design}
\newacronym{PIC}{PIC}{Particle In Cell}
\newacronym{CST}{CST}{Computer Simulation Technology}
\newacronym{SE}{SE}{Secondary Electron}
\newacronym{FC}{FC}{Faraday Cup}
\newacronym{BLM}{BLM}{Beam Loss Monitor}

\newglossaryentry{DUT}
{
	type=\acronymtype,
	name={DUT},
	description={Device Under Test},
	first={Device Under Test (DUT)},
	plural={\glsentrytext{DUT}s},
	descriptionplural={Devices Under Test},
	firstplural={Devices Under Test (DUTs)}
} 

\makeglossaries
\journal{Elsevier}
\DeclareSIUnit{\NEQ}{\text{$\mathrm{n_{eq}}$}}
\begin{document}

	\begin{frontmatter}
		
		\title{A Beam-Driven Proton Irradiation Setup for Precision Radiation Damage Tests of Silicon Detectors}
		
		\author[PI]{Pascal Wolf\corref{cor1}}
		\ead{wolf@physik.uni-bonn.de}
		\cortext[cor1]{Corresponding author}
		
		\author[HISKP]{Dennis Sauerland}
		
		\author[HISKP]{Reinhard Beck}
		
		\author[PI]{Jochen Dingfelder}
		
		\address[PI]{\gls{PI}, Rheinische Friedrich-Wilhelms-Universit{\"a}t Bonn, Nu{\ss}allee 12, 53115 Bonn, Germany}
		\address[HISKP]{\gls{HISKP}, Rheinische Friedrich-Wilhelms-Universit{\"a}t Bonn, Nu{\ss}allee 14-16, 53115 Bonn, Germany}
		
		\begin{abstract}

			A proton irradiation site for silicon detectors has been developed and commissioned at the Bonn Isochronous Cyclotron.
			The accelerator provides \qty{14}{\mega\electronvolt} proton beams of up to \qty{1}{\micro\ampere} at beam widths of a few \unit{\milli\meter} to the setup.
			\Glspl{DUT} are irradiated inside a cooled, thermally-insulated box at $\le\qty{-20}{\degreeCelsius}$, while being moved through the beam in a row-based scan pattern to achieve uniform fluence distributions.
			Custom-made diagnostics allow for beam-based, on- and offline dosimetry, enabling a beam-driven irradiation routine which produces uniform fluence distributions with standard deviations $\ll\qty{1}{\percent}$.
			Dedicated irradiations of thin titanium foils are performed to compare the commonly-used dosimetry via metallic foil activation to the beam-based approach. Within the error margins, both methods are in agreement, whereas the beam-based technique yields lower uncertainties of typically $\le\qty{2}{\percent}$.
			Simulations indicate a reduction of the initial proton energy to \qty{12.28\pm0.06}{\mega\electronvolt} on the \gls{DUT}.
			Characterization of six \qty{150}{\micro\meter}-thin, passive LFoundry sensors before and after irradiation yield a proton hardness factor 
			of $\kappa_\mathrm{p}=3.71\pm0.11$, which is in agreement with expectations, allowing to irradiate up to \qty{e16}{\NEQ\per\centi\meter\squared} within a few hours.
		\end{abstract}
		
	
	\end{frontmatter}
%
%
\section{Irradiation Site}
\label{sec:irradiation_site}
%
%
\subsection{Setup}
\label{subsec:irrad_setup}
\begin{figure}
	\centering
	\includegraphics[width=0.5\textwidth]{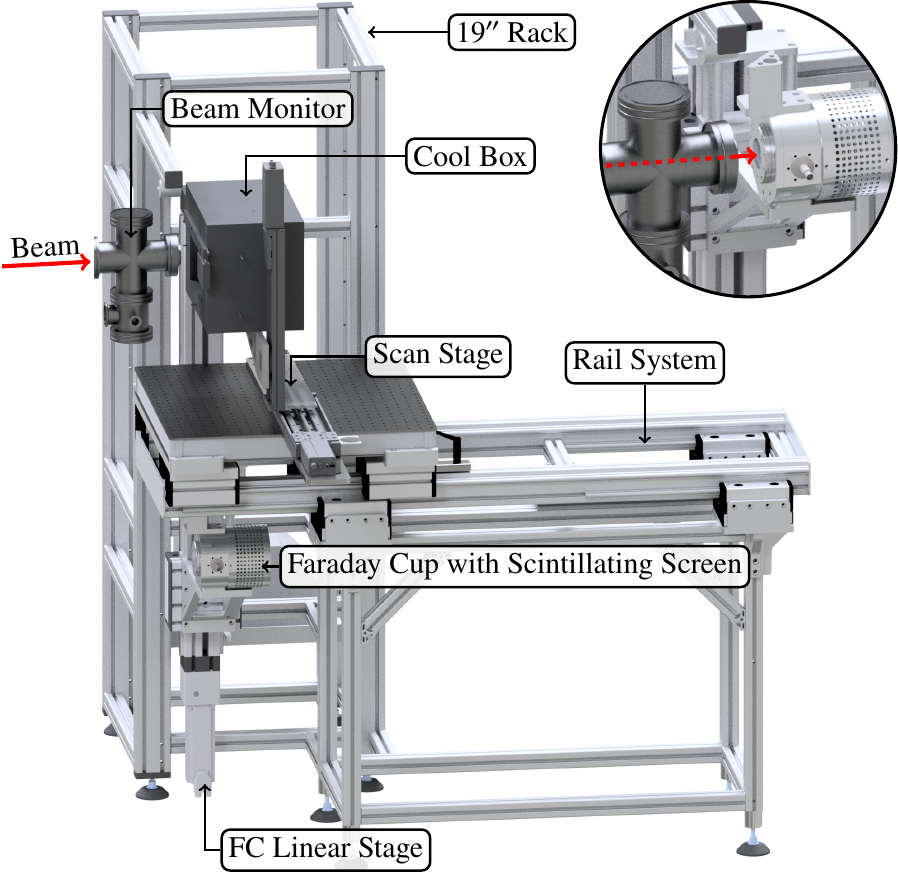}
	\caption[Setup]{\gls{CAD} render of the setup, adapted from \cite{ipac22}. Upper right shows the calibration configuration.}
	\label{cad:setup}
\end{figure}

The irradiation site is located at the Bonn Isochronous Cyclotron which typically provides light ions up to alpha particles between \qtyrange{7}{14}{\mega\electronvolt} per nucleon.
Commonly, the accelerator delivers a \qty{13.6}{\mega\electronvolt} ($\approx$ \qty{12.3}{\mega\electronvolt} on \gls{DUT}) proton beam with currents between \qty{20}{\nano\ampere} to \qty{1}{\micro\ampere} and widths of a few \unit{\milli\meter} \gls{FWHM} to the site.
The setup is depicted in \cref{cad:setup}.
The beam passes through a calibrated beam monitor into a thermally-insulated cool box, mounted on a \textit{scan stage} consisting of two linear stages.
The scan stage is installed on a retractable table, allowing a \gls{FC} on a linear stage to be driven in front of the beam monitor for calibration, replacing the cool box.
\subsection{Devices Under Test}
\label{subsec:irrad_duts}
\begin{figure}
	\centering
	\includegraphics[width=0.5\textwidth]{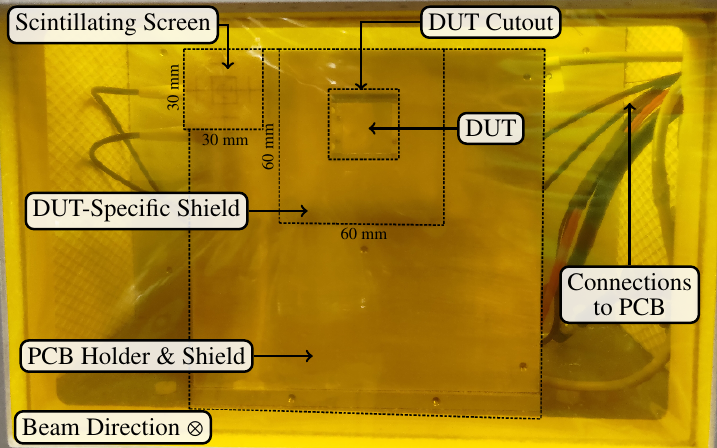}
	\caption[Cool box]{Front view of cool box with an installed carrier \gls{PCB}, exposing only the \gls{DUT} to the beam.}
	\label{pic:cool_box}
\end{figure}

Typical \glspl{DUT} arrive at the site on a carrier \gls{PCB} or \gls{SCC}. They are mounted behind a \qty{6}{\milli\meter}-thick Al-shield inside the cool box.
A front view of the box through its Kapton entrance window is shown in \cref{pic:cool_box}.
The shield is composed of a generic sample holder as well as a \gls{DUT}-specific, in-house manufactured cut-out, exposing only the silicon to the beam.
On the top-left of the shield, a scintillating screen is located, enabling beam-based alignment of the setup and serving as the origin position of the irradiation routine.
A feed-through allows to connect power and readout cables to the \gls{DUT} during irradiation.
Other \glspl{DUT} not situated on carrier \glspl{PCB}, such as bare sensors and diodes, are irradiated on a dedicated carrier plate.
During irradiation, the box is continuously flushed with nitrogen gas, cooled by a liquid nitrogen heat exchanger, maintaining temperatures of $\le\qty{-20}{\degreeCelsius}$ to prevent annealing and a dry atmosphere to avoid condensation.
\section{Beam Monitoring}
\label{sec:beam_monitoring}
%
%
\subsection{Principle}
\label{subsec:see_principle}
The beam diagnostics are based on the effect of \gls{SEE} on the surface of materials penetrated by fast ions as described in \cite{GARNIR1982187}.
The ratio of emitted electrons per initial ion is called \gls{SEY} $\gamma$ and can be defined using the \gls{SEE} as well as ion beam current as
\begin{align}
	\label{eq:sey}
	\gamma = \frac{I_\mathrm{SEE}}{I_\mathrm{beam}} \cdot z_\mathrm{ion}\,,
\end{align}
where $z_\mathrm{ion}$ is the number of elementary charges $q_e$ carried by the ion.
Generally, $I_\mathrm{SEE}$ depends on a variety of parameters such as pressure, target temperature, ion type and energy as well as beam intensity \cite{GARNIR1982187}.
Assuming a monoenergetic proton beam inside a vacuum, $\gamma$ is approximately constant over orders of magnitude of beam intensity and \cref{eq:sey} yields a constant.
Allowing the beam to penetrate a thin foil enables online beam monitoring by one-time determination of $\gamma$ and subsequent, continuous measurement of $I_\mathrm{SEE}$.
\begin{figure}
	\centering
	\includegraphics[width=0.5\textwidth]{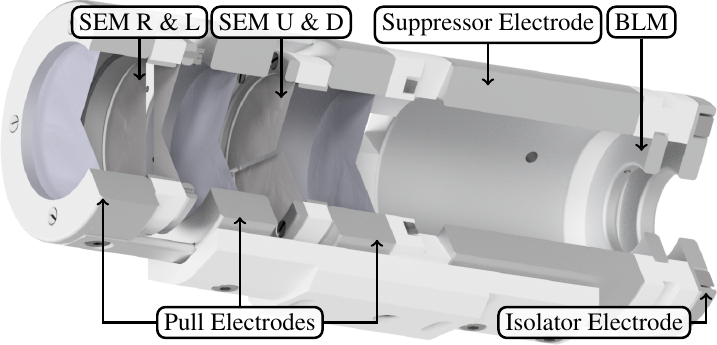}
	\caption[Beam monitor]{\gls{CAD} render of the custom-made beam monitor from \cite{ipac22}. The beam passes from left to right.}
	\label{cad:beam_monitor}
\end{figure}
\subsection{Custom Monitor}
\label{subsec:custom_monitor}
\begin{figure}
	\centering
	\includegraphics[width=0.5\textwidth]{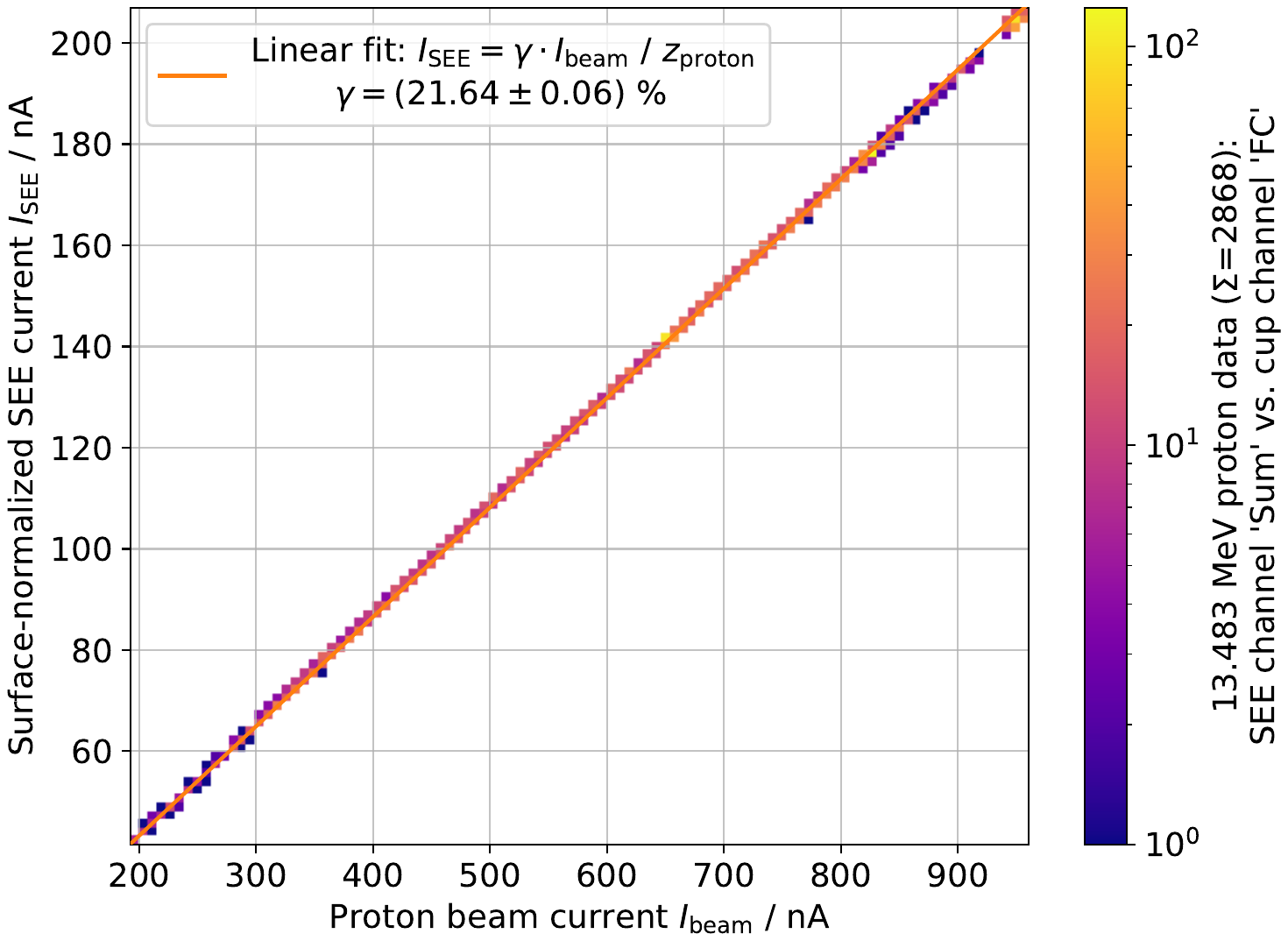}
	\caption[Calibration]{Beam monitor calibration for \qty{13.483}{\mega\electronvolt} protons.}
	\label{plt:calibration}
\end{figure}

Building on this principle, a custom beam monitor concept was developed and is depicted in \cref{cad:beam_monitor}.
It consists of a \gls{SEM} as well as a \gls{BLM} module for beam current and position as well as beam loss monitoring.
The \gls{SEM} is composed of two, carbon-coated ($\ge\qty{70}{\nano\meter}$, to anticipate carbon build-up due to irradiation \cite{Peterson1977} and subsequent change of \gls{SEY}), \qty{4.5}{\micro\meter} Al-foil pairs, segmented in the vertical (SEM L+R) and horizontal plane (SEM U+D).
Al-foil pull electrodes are placed in between at \qty{+100}{\volt}, removing \glspl{SE} from the \gls{SEM} foils.
Subsequently, the individual \gls{SEE} currents of each respective \gls{SEM} foil $I_\mathrm{SEE}\left(\mathrm{L|R|U|D}\right)$ can be measured.
The \gls{BLM} module consists of a \qty{3}{\milli\meter}-thick Al-iris enclosed by a suppressor as well as isolator electrode at \qty{-100}{\volt} and functions as an internal \gls{FC}, allowing to detect beam cut-off.
The \gls{SEM} currents in combination with a beam current measurement via an external \gls{FC}, as shown in \cref{cad:setup}, can be used to calibrate the beam monitor according to \cref{eq:sey}.
This is depicted for protons in \cref{plt:calibration}, where $\gamma=(21.64\pm0.06)\%$ over approximately an order of magnitude of beam current.
The current signals of the beam diagnostics are converted to voltages using a custom, analog readout board. It features a transimpedance amplifier chain for each signal channel with variable input current scales, allowing to adapt to different signal amplitudes. The resulting voltages are mapped to $\pm 5\si{\volt}$ and digitized using an \gls{ADC} board. The relative uncertainty of the readout chain is in the order of \qty{2}{\percent}. More detailed description of the diagnostics can be found in \cite{ipac22,cyc22}.
\section{Irradiation Procedure}
\label{sec:irradiation_procedure}
%
%
\subsection{Preparation}
\label{subsec:scan_setup}
A beam monitor calibration is performed prior to each irradiation to compensate systematic errors due to run-by-run variations in ion energy and environmental parameters. Moreover, a maximum estimate of the horizontal and vertical beam \gls{FWHM} is made from visual inspection on the scintillating screen on the \gls{FC} with a camera, to account for them in the irradiation routine.
The \gls{DUT} is installed in the cool box and cooled to $\le\qty{-20}{\degreeCelsius}$.
Subsequently, the setup is aligned by simultaneously centering the beam spot in the beam monitor as well as on the scintillation screen of the cool box, using a camera, defining the origin position for the irradiation.
\subsection{Routine}
\label{subsec:scan_scheme}
\begin{figure}
	\centering
	\includegraphics[width=0.5\textwidth]{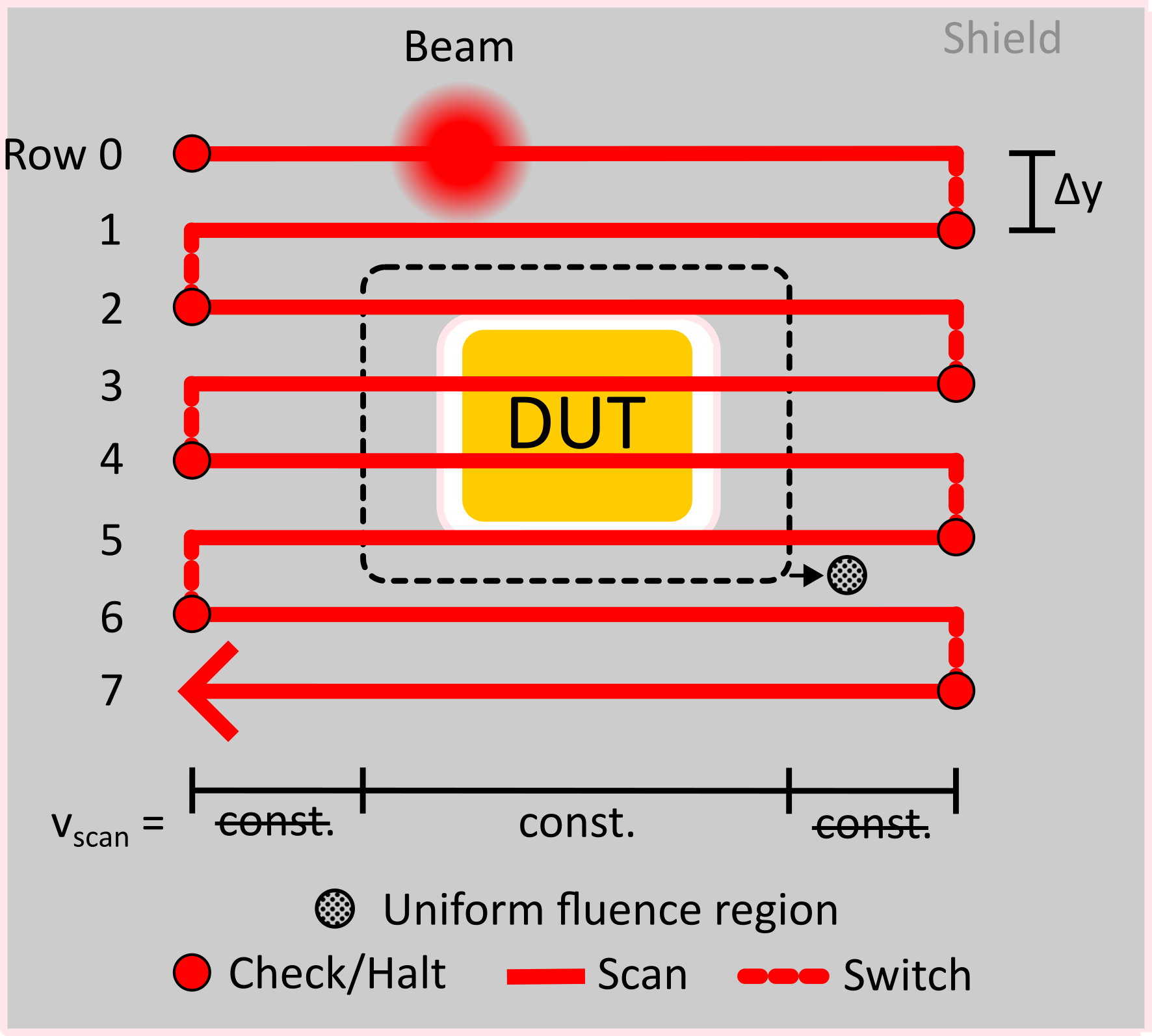}
	\caption[Irradiation routine]{Irradiation pattern. Beam parameters are checked prior to scanning each row}
	\label{schm:scan}
\end{figure}

The irradiation routine is shown schematically in \cref{schm:scan}.
In reference to the origin position, a grid with equidistantly-spaced rows is generated, using the \gls{DUT}'s position and cross section, on which it is moved through the stationary beam.
The assembled grid covers an area larger than the \gls{DUT} cross section to include margins considering the beam \gls{FWHM} as well as the required acceleration distance to reach the scan velocity $v_\mathrm{scan}$.
Choosing a row separation $\Delta y \ll \text{FWHM}$ and ensuring a constant velocity when traversing the \gls{DUT}, results in the integration of the beam profile along the respective dimensions, producing a uniform fluence distribution around the \gls{DUT} as indicated in \cref{schm:scan}.
The traversal of all rows of the grid constitutes a \textit{complete scan}.
After completing a scan, the procedure is repeated, starting from the previously-scanned row, until the target fluence is applied.
\subsection{Beam-Based Irradiation}
\label{subsec:scan_interaction}
\begin{figure}
	\centering
	\includegraphics[width=0.5\textwidth]{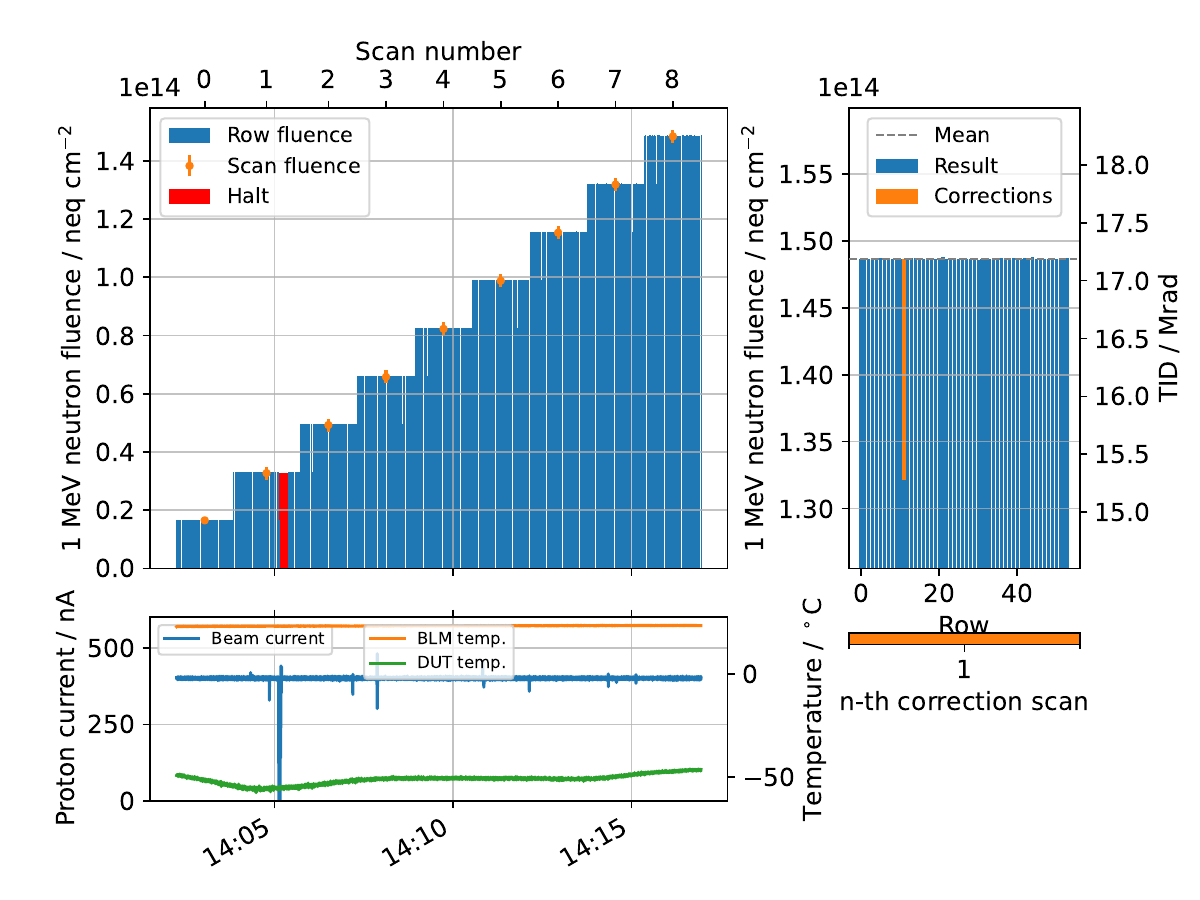}
	\caption[Irradiation overview]{Overview of fluence per row and scan (top), beam current and temperatures (bottom) and the resulting fluence distribution (right).}
	\label{plt:scan}
\end{figure}

Continuous monitoring of the beam parameters during the irradiation procedure enables to define beam-based (and other) \textit{events}, to which the irradiation routine can react.
An event is either active or inactive and is checked for at the beginning of each row as shown in \cref{schm:scan}.
In the case of one or more active events, the scan routine is halted while the beam remains on the shield, not irradiating the \gls{DUT}, until all events are inactive.
Typical events check for stable as well as sufficient beam current, a centered beam position but also adequately-low \gls{DUT} temperature.
Events becoming active while scanning a row are recorded and their effect on the fluence distribution can be accounted or even corrected for as shown in \cref{plt:scan}.
Here, due to a drop in beam current while traversing the \gls{DUT}, a row was re-scanned post-irradiation with adjusted parameters to level the fluence distribution.
This minimizes exposing the \gls{DUT} to deficient beam conditions, maximizing the uniformity of the applied fluence.
\section{Dosimetry}
\label{sec:dosimetry}
%
%
\subsection{Foil Activation Method}
\label{subsec:foil_activation}
Conventionally, the determination of the primary particle fluence is performed via activation of metallic foils \cite{dierlamm10}.
Here, depending on the ion species as well as energy, metals with an adequate isotope X of known production cross section $\Omega_\mathrm{X}$ are selected and irradiated alongside the \gls{DUT} within the uniform area indicated in \cref{schm:scan}.
After measurement of the resulting activity $A_\mathrm{X}$, the particle fluence is a function of
\begin{align}
	\Phi \left(\Omega_\mathrm{X}, A_\mathrm{X},  M_\mathrm{foil}, m_\mathrm{X}, \lambda_\mathrm{X}\right)\,,
\end{align}
where $M_\mathrm{foil}$ is the foil mass, $m_\mathrm{X}$ the isotopes molar mass and $\lambda_\mathrm{X}$ its decay constant.
The obtained fluence is a scalar value and the relative uncertainty typically in the order of $\Delta \Phi / \Phi \approx \qty{10}{\percent}$ \cite{dierlamm10}.
\subsection{Beam-Based Methods}
\label{subsec:beam_data}
\begin{figure}
	\centering
	\includegraphics[width=0.5\textwidth]{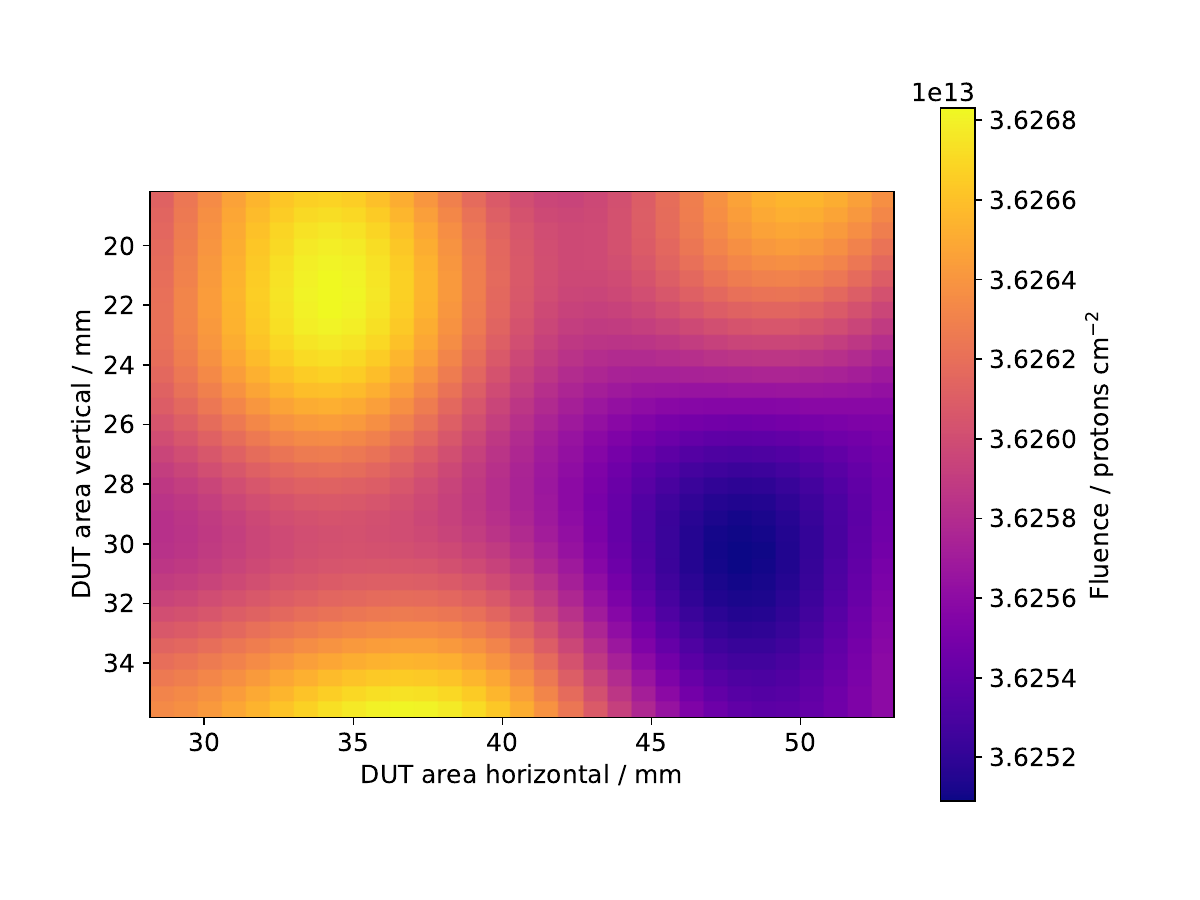}
	\caption[2D fluence distribution]{2D fluence distribution on the \gls{DUT} area, generated from the data recorded during irradiation. With a fluence of $\num{3.6e13}\pm\num{4.4e9}$ \unit{\text{p}\per\centi\metre\squared}, the relative standard deviation is $\ll\qty{1}{\percent}$.}
	\label{plt:fluence}
\end{figure}

The continuous beam parameter monitoring allows for a purely beam-based, on- and offline dosimetry with a relative uncertainty of typically \qty{2}{\percent}, dominated by the beam current measurement. 
Following the procedure shown in \cref{schm:scan}, the fluence applied within the uniform area per \textit{complete scan} can be geometrically approximated as \cite{dierlamm10}
\begin{align}
	\label{eq:fluence_det_formula}
	\Phi = \frac{I_\mathrm{beam}}{z_\mathrm{ion} \cdot q_e \cdot v_\mathrm{scan}\cdot \Delta y}\,,
\end{align}
where $I_\mathrm{beam}$ is the ion beam current and $\Delta y$ is the row separation.
This allows for online monitoring the fluence with row resolution, as shown in \cref{plt:scan}.
Here, corrections can be applied after identifying rows with insufficient fluence and re-irradiating them with adjusted parameters.
Furthermore, the online fluence monitoring enables to pause the irradiation routine at any given fluence level to perform \gls{DUT} measurements as well as applying custom fluence profiles.\\\newline
Additionally, the data recorded during irradiation allows to generate a two-dimensional distribution of the particle fluence on the scan pattern as well as \gls{DUT} area after irradiation.
Here, the continuous monitoring of beam and setup parameters enables to interpolate the beam current along the paths of the irradiation routine.
Assuming a Gaussian beam profile using the maximum estimations for the widths from the alignment process in \cref{subsec:irrad_setup}, a fluence distribution can be calculated for every point on the area by applying a Gaussian kernel.
The resulting distribution on the \gls{DUT} area is shown in \cref{plt:fluence}.
With a relative standard deviation of $\ll\qty{1}{\percent}$, the fluence can be assumed uniform within the uncertainties.
\subsection{Comparison}
\label{subsec:comparison}
\begin{figure}
	\centering
	\includegraphics[width=0.5\textwidth]{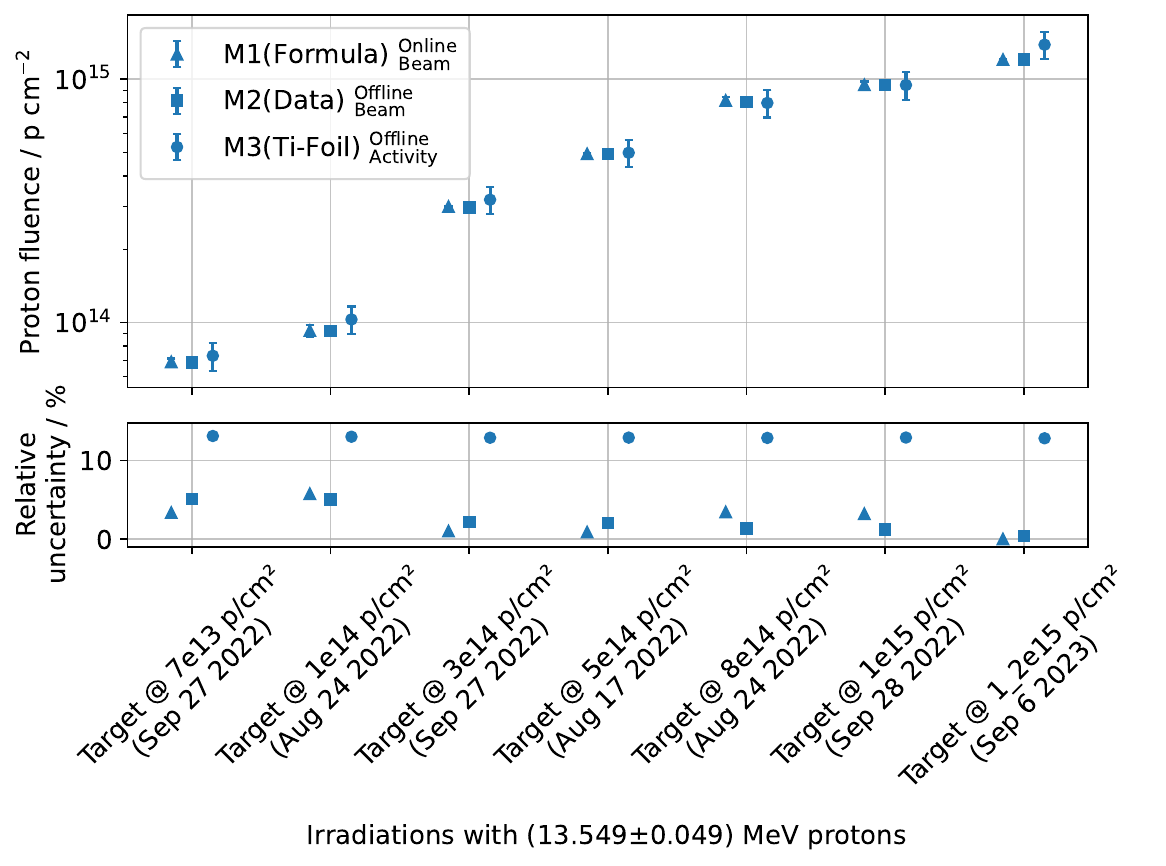}
	\caption[Dosimetry comparison]{Comparison of dosimetry via beam-based methods (M1 \& M2) and via foil activation (M3).}
	\label{plt:dosimetry_comp}
\end{figure}

To verify that the beam-based dosimetry methods via \cref{eq:fluence_det_formula} (M1) and irradiation data analysis (M2) as well as the dosimetry via foil activation (M3) yield comparable results, seven titanium foils were irradiated with \qty{13.55\pm0.05}{\mega\electronvolt} protons, corresponding to \qty{12.22\pm0.05}{\mega\electronvolt} on-foil, to fluences between \qtyrange{7e13}{12e14}{\text{p}\per\centi\metre\squared}.
Here, the vanadium isotope $\mathrm{^{48}V}$ is produced, via $\mathrm{^{48}Ti\ \overset{(p,n)}{\rightarrow}\ ^{48}V}$, with a cross section of $\mathrm{\Omega_{^{48}V}}=\qty{412\pm47}{\milli\barn}$ for \qty{12.5\pm0.2}{\mega\electronvolt} protons \cite{natTitoV48}.
The resulting proton fluences as well as their relative uncertainties are shown in \cref{plt:dosimetry_comp}.
All methods yield consistent results whereas the beam-based approaches produce significantly lower, relative uncertainties of $\le\qty{5}{\percent}$ and additionally provide spatial resolution.
\section{Proton Hardness Factor}
\label{sec:hardness_factor}
%
%
\subsection{Theory}
\label{subsec:niel_theory}
\begin{figure}
	\centering
	\includegraphics[width=0.5\textwidth]{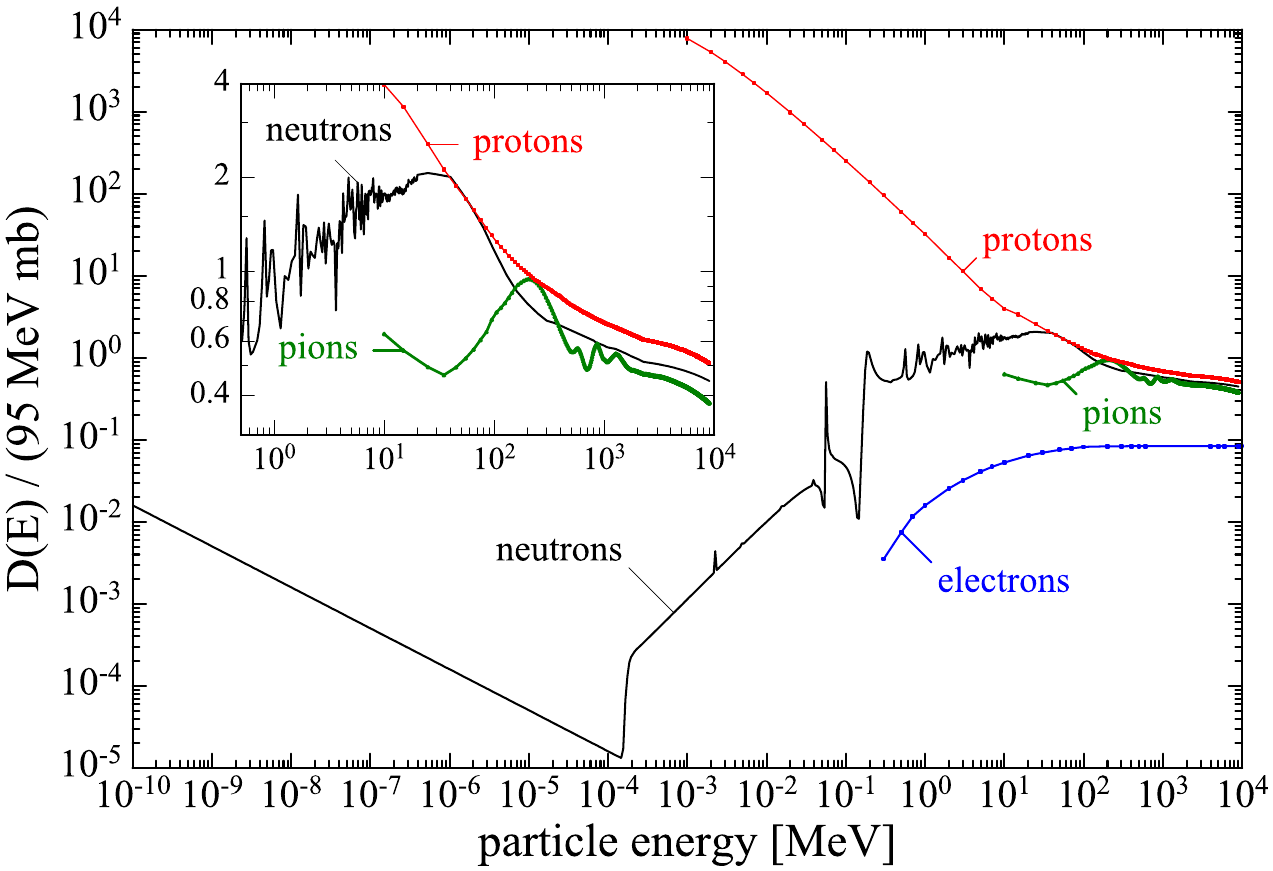}
	\caption[Particle \gls{NIEL}]{\gls{NIEL} damage versus energy for various particles from \cite{MOLL18}}
	\label{plt:niel}
\end{figure}

\begin{figure}
	\centering
	\includegraphics[width=0.5\textwidth]{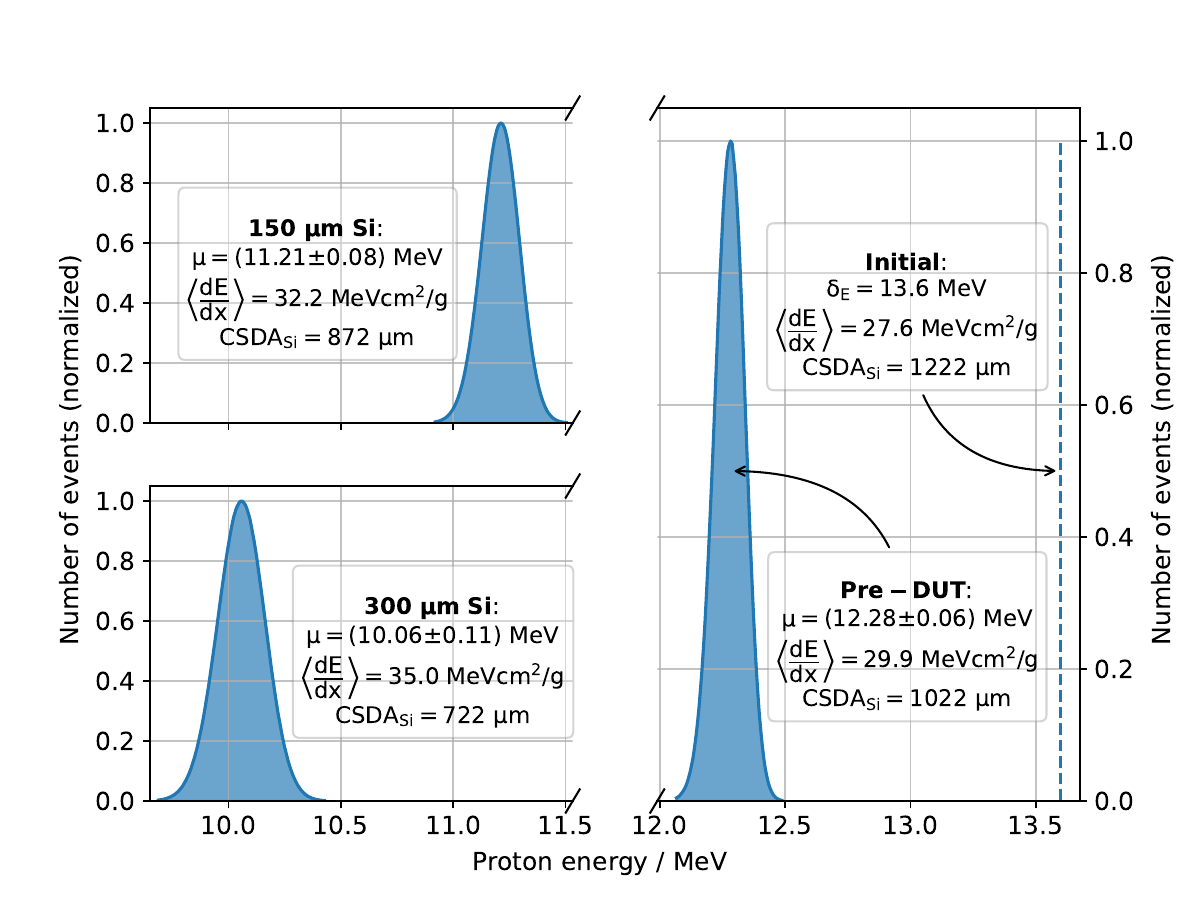}
	\caption[Proton energy simulation]{\textit{GEANT4} simulation of initial, pre- and post-\gls{DUT} proton energy distributions for two silicon thicknesses. 
	}
	\label{plt:proton_energy_sim}
\end{figure}

The hardness factor $\kappa$ is a scaling factor, representing a particles \gls{NIEL} damage, normalized to \qty{1}{\mega\electronvolt} neutron equivalents $\unit{\NEQ}$.
It is given in \unit{\mega\electronvolt\milli\barn} and depends on the particle species as well as energy, as depicted in \cref{plt:niel}.
For protons of \qty{13.6}{\mega\electronvolt}, \textit{GEANT4} \cite{AGOSTINELLI2003250} simulations displayed in \cref{plt:proton_energy_sim} yield an energy degradation to \qty{12.28\pm0.06}{\mega\electronvolt} when reaching the \gls{DUT}'s surface and \qty{10.06\pm0.11}{\mega\electronvolt} after \qty{300}{\micro\meter} silicon. For these energies, a hardness factor of $\kappa_\mathrm{p} \approx 4$ is expected.
The \gls{NIEL} hypothesis states that sensor properties such as leakage current scale linearly with the \gls{NIEL} damage and therefore the particle fluence \cite{MOLL18}.
Here, the increase in leakage current $\Delta I_\mathrm{leak}$ per depleted volume $V$ is directly proportional to the fluence $\Phi$
\begin{align}
\label{eq:niel_saling}
\frac{\Delta I_\mathrm{leak}}{V} = \alpha \cdot \Phi\,,
\end{align}
where $\alpha$ is the \textit{current-related damage factor} which itself is a function of annealing time and temperature.
Conventionally, annealing is performed for \qty{80}{\minute} at \qty{60}{\degreeCelsius} and leakage current measurements are performed at or scaled to \qty{20}{\degreeCelsius} via \cite{Chilin13}
\begin{align}
\label{eq:leakage_scaling}
I_\mathrm{leak}\left(T\right)\propto T^2 \cdot \exp\left(-\frac{E_\mathrm{eff}}{2k_BT}\right)
\end{align}
with $E_\mathrm{eff}=\qty{1.214\pm0.014}{\electronvolt}$.
Given the above conventions, $\alpha_{\unit{\NEQ}}=\qty{3.99\pm0.03e-17}{\ampere\per\centi\meter}$ for \qty{1}{\mega\electronvolt} neutrons \cite{MOLL18}.
With knowledge of the current-related damage factor $\alpha_\mathrm{x}$ for a given particle x, its hardness factor can be defined as
\begin{align}
\label{eq:hardness_factor}
\kappa_\mathrm{x} = \frac{\alpha_\mathrm{x}}{\alpha_{\unit{\NEQ}}}\,.
\end{align}
\subsection{Measurement}
\label{subsec:kappa_measurement}
\begin{figure}[h]
	\centering
	\includegraphics[width=0.5\textwidth]{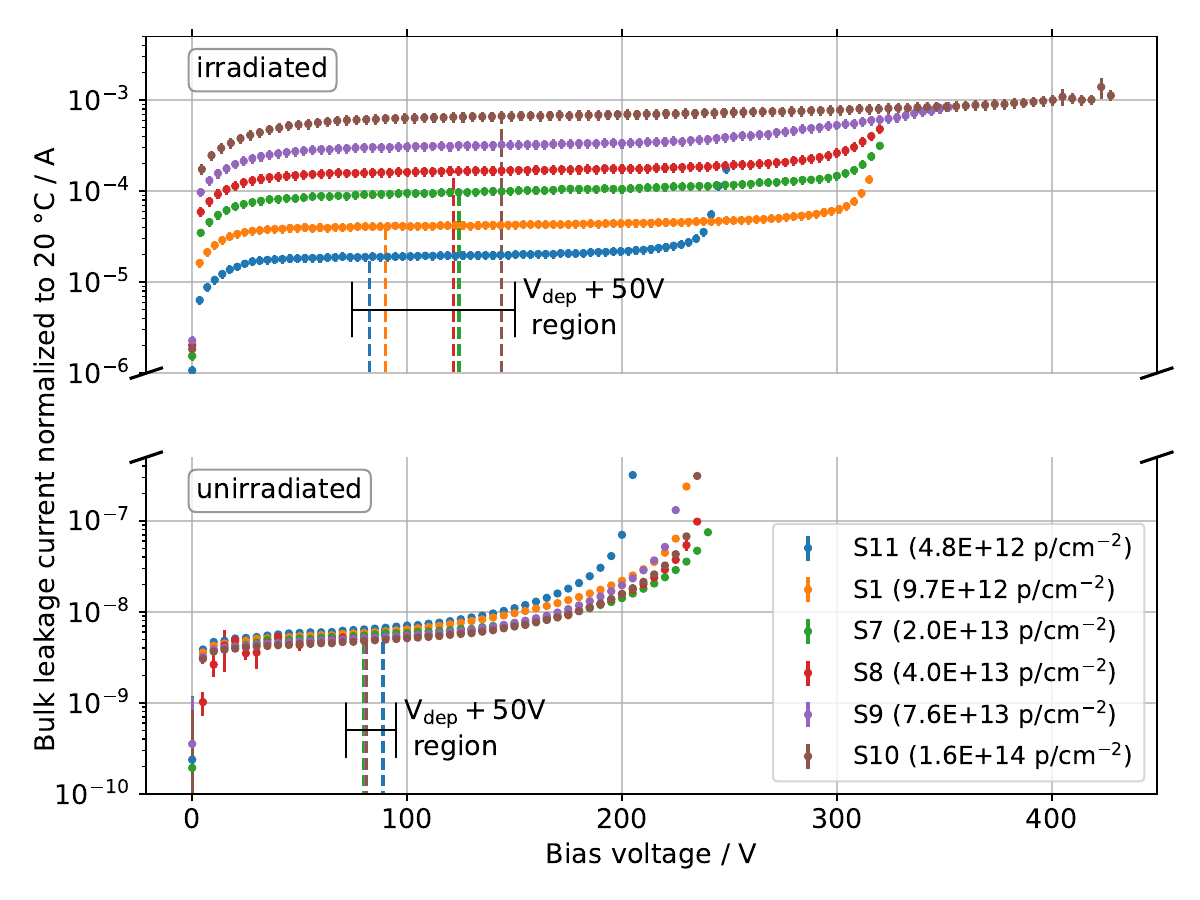}
	\caption[IV curves of sensors]{IV curves scaled to  \qty{20}{\degreeCelsius} of \qty{150}{\micro\meter} thick sensors before and after irradiation. The full depletion voltage $\mathrm{V_{dep}}$ was determined by CV characterization as described in \cite{maegdefessel22}.}
	\label{plt:iv_sensors}
\end{figure}

\begin{figure}[h]
	\centering
	\includegraphics[width=0.5\textwidth]{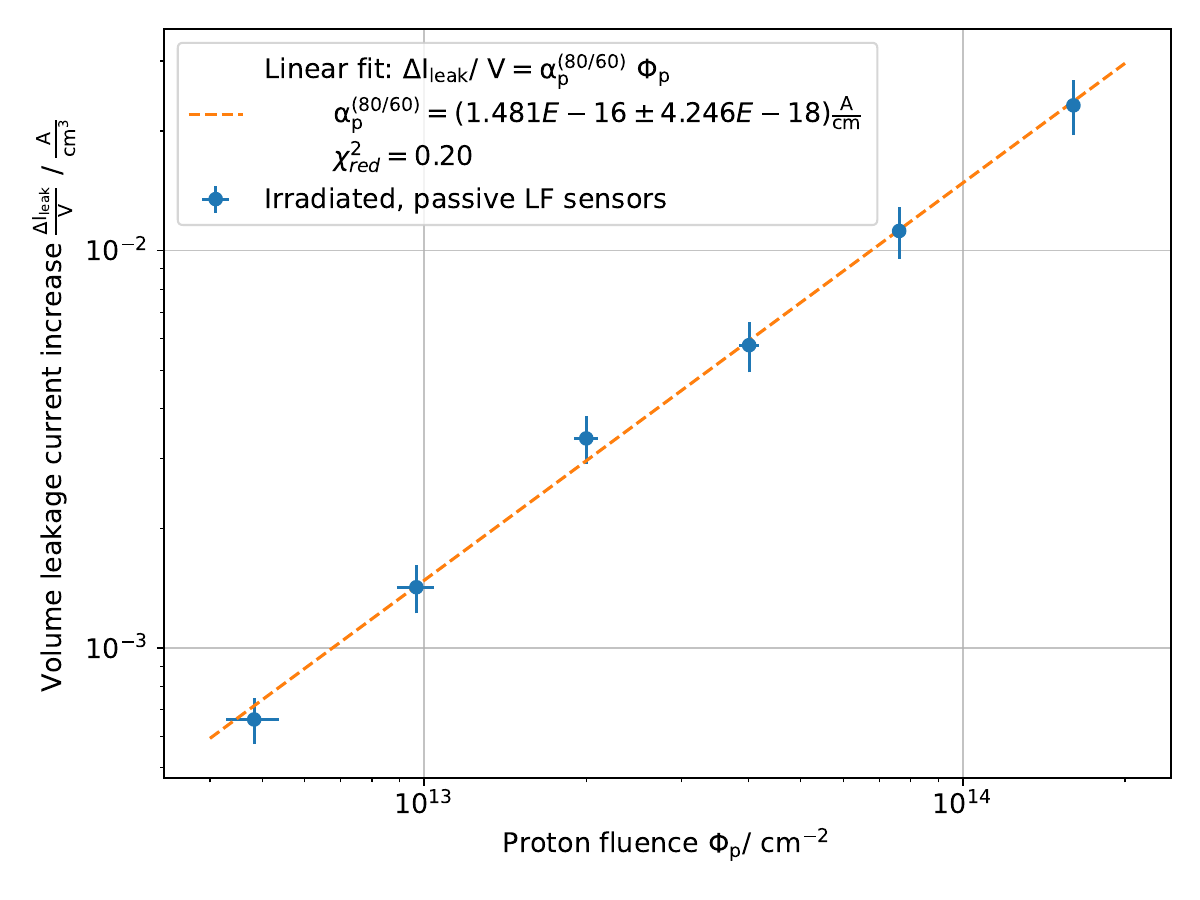}
	\caption[Hardness factor]{Leakage current increase per depleted volume versus applied proton fluence.}
	\label{plt:hardness_factor}
\end{figure}

Using \cref{eq:hardness_factor,eq:niel_saling}, the proton hardness factor is obtained by irradiation of sensors to different fluences as well as measurement of the leakage current at full depletion.
Therefore, six \qty{150}{\micro\meter}-thick, $\left(\qty{1.92}{}\times\qty{0.96}{}\right)\unit{\centi\meter\squared}$, passive sensors from LFoundry were irradiated with \qty{13.52\pm0.04}{\mega\electronvolt} protons to different fluences between \qtyrange{5e12}{16e13}{\text{p}\per\centi\metre\squared} and electrically characterized.
After irradiation, the sensors were annealed for \qty{80}{\minute} at \qty{60}{\degreeCelsius} and their leakage currents measured in a temperature-controlled environment between \qtyrange{-10}{-25}{\degreeCelsius} with an error of \qty{1}{\degreeCelsius}.
Their IV behavior, scaled to \qty{20}{\degreeCelsius} via \cref{eq:leakage_scaling}, pre- and post-irradiation is depicted in \cref{plt:iv_sensors}.
Using the method described in \cite{maegdefessel22}, the full depletion voltage $V_\mathrm{dep}$ after irradiation for each sensor is determined via CV measurements at different frequencies.
The expected increase of leakage current with fluence can be observed.
The leakage current is evaluated at $V_\mathrm{dep}+\qty{50}{\volt}$ to ensure full depletion where an uncertainty of \qty{10}{\micro\meter} is assumed on the sensor thickness to account for processing variances.
To obtain $\alpha_p$, the increase in leakage current per depleted volume is plotted versus the respective proton fluence, as shown in \cref{plt:hardness_factor}.
A linear fit according to \cref{eq:niel_saling} was performed and $\alpha_\mathrm{p}=\qty{1.48\pm0.04e-16}{\ampere\per\centi\meter}$ was extracted which yields a proton hardness factor of
\begin{align*}
\kappa_\mathrm{p} = \frac{\alpha_\mathrm{p}}{\alpha_{\unit{\NEQ}}}=\num{3.71\pm0.11}\,.
\end{align*}
This is in agreement with the data shown in \cref{plt:niel} for the expected \qty{12.3}{\mega\electronvolt} protons on the \gls{DUT}, enabling irradiation of up \qty{e16}{\NEQ\per\centi\meter\squared} within a few hours.
\section{Limitations}
\label{sec:limitations}
The low proton energy of \qty{12.3}{\mega\electronvolt} on the \gls{DUT} constrains the \gls{NIEL} scaling accuracy for thick silicon devices.
As indicated in \cref{plt:niel}, the hardness factor strongly-depends on the proton energy in this regime.
The energy loss due to ionization while traversing the \gls{DUT} increases the hardness factor gradually.
Here, for a silicon thickness of $\qty{300}{\micro\meter}$, an energy loss of \qty{2.2}{\mega\electronvolt} (see \cref{plt:proton_energy_sim}) corresponds to an expected increase of the hardness factor of approximately \qty{10}{\percent} between entry and exit.
Therefore, the stated proton hardness factor of $\kappa_\mathrm{p}=\num{3.71\pm0.11}$ is only accurate for \glspl{DUT} $\le\qty{150}{\micro\meter}$ silicon and generally a thickness $\le\qty{300}{\micro\meter}$ is preferable.
For thicker devices up to $\qty{300}{\micro\meter}$, an increased hardness factor as well as uncertainty of $\kappa_\mathrm{p}=\num{4.0\pm0.4}$ is assumed, in accordance to the expected increase due to energy loss by ionization.
\section{Conclusion}
\label{sec:conclusion}
In this work, a modern proton irradiation site at the \gls{BIC} of Bonn University is described. \Glspl{DUT} are irradiated with \qty{13.6}{\mega\electronvolt} protons inside a cool box at $\le\qty{-20}{\degreeCelsius}$ which is moved trough the stationary beam along a scan grid. Custom beam diagnostics facilitate online beam parameter monitoring at the extraction to the setup, enabling a beam-driven irradiation routine as well as a purely beam-based on- and offline dosimetry. Test irradiations of titanium foils verify that the beam-based and standard dosimetry via foil activation yield comparable results whereas the beam-based methods provide a lower uncertainty. Performing electrical characterization of thin sensors before and after irradiation allows to extract a proton hardness of $\kappa_\mathrm{p}=\num{3.71\pm0.11}$, in agreement with expectations. Due to the low proton energy at the \gls{BIC}, \glspl{DUT} with a thickness $\le\qty{300}{\micro\meter}$ silicon are preferable to ensure accurate \gls{NIEL} scaling.
%
\bibliography{submission.bib}
\end{document}